\documentclass[pra,10pt,showpacs,amsmath,twocolumn,floatfix,eqsecnum]{revtex4}
\usepackage{amsmath}
\usepackage{amsfonts}
\usepackage{graphicx}

\newcommand{\parti}[2]{\frac{\partial #1}{\partial #2}}
\newcommand{\partit}[2]{\frac{\partial^2 #1}{\partial #2^2}}

\newcommand{\ket}[1]{|#1\rangle}

\newcommand{\bra}[1]{\langle#1|}

\newcommand{\avg}[1]{\langle#1\rangle}
\newcommand{\Avg}[1]{\left\langle#1\right\rangle}

\newcommand{\bk}[1]{\left(#1\right)}
\newcommand{\Bk}[1]{\left[#1\right]}
\newcommand{\BK}[1]{\left\{#1\right\}}

\newcommand{\trace}{\operatorname{tr}}

\begin{document}
\title{Optimal waveform estimation for classical and quantum systems
  via time-symmetric smoothing. II. Applications to atomic
  magnetometry and Hardy's paradox}

\author{Mankei Tsang}

\email{mankei@mit.edu}

\affiliation{Research Laboratory of Electronics,
Massachusetts Institute of Technology, Cambridge, Massachusetts
02139, USA}






\date{\today}

\begin{abstract}
  The quantum smoothing theory [Tsang, \prl \textbf{102}, 250403
  (2009); Phys.\ Rev.\ A, in press (e-print arXiv:0906.4133)] is
  extended to account for discrete jumps in the classical random
  process to be estimated, discrete variables in the quantum system,
  such as spin, angular momentum, and photon number, and Poissonian
  measurements, such as photon counting. The extended theory is used
  to model atomic magnetometers and study Hardy's paradox in phase
  space.  In the phase-space picture of Hardy's proposed experiment,
  the negativity of the predictive Wigner distribution is identified
  as the culprit of the disagreement between classical reasoning and
  quantum mechanics.
\end{abstract}

\maketitle
\section{Introduction}
In previous papers \cite{tsang_prl, tsang_pra}, I have proposed a
quantum smoothing theory, which can be used to optimally estimate
classical signals coupled to quantum sensors under continuous
measurements, such as gravitational wave detectors and atomic
magnetometers. Smoothing can be significantly more accurate than
current quantum filtering methods
\cite{belavkin,carmichael,gardiner_zoller,mabuchi,berry,thomsen,geremia,stockton,bouten}
when the classical signal is a stochastic process and delay is
permitted in the estimation. While Refs.~\cite{tsang_prl,tsang_pra}
focus on diffusive classical random processes, quantum systems with
continuous degrees of freedom, and Gaussian measurements, the aim of
this paper is to extend the theory to account for discrete variables
in the systems and the measurements. In particular, I shall consider
discrete jumps in the classical random process, discrete variables in
the quantum system, such as spin, angular momentum, and photon number,
and Poissonian measurements, such as photon counting. Such extensions
are especially important for the modeling of atomic magnetometry
\cite{thomsen,geremia,stockton,budker,kuzmich,petersen,bouten}.

In the case of atomic magnetometry, the importance of estimation delay
was discovered by Petersen and M{\o}lmer \cite{petersen}, who found
that the estimation of a fluctuating magnetic field modeled as an
Ornstein-Uhlenbeck process becomes more accurate when the estimation
is delayed and observations at later times are taken into account. I
shall generalize their results using the quantum smoothing theory,
derive the optimal strategy of delayed estimation for atomic
magnetometry, and discuss practical methods of implementing the
strategy.

A different kind of estimation problem comes up in Hardy's paradox
\cite{hardy}, in which one estimates the positions of an electron and
a positron in interferometers based on posterior measurement outcomes
and obtains paradoxical results. I shall demonstrate that the salient
features of the paradox can be reproduced mathematically using the
quantum smoothing theory in discrete phase space, which is arguably
the most natural way of modeling classical properties of quantum
objects.  It is shown that the negativity of the predictive Wigner
distribution can be regarded as the culprit of the disagreement
between classical reasoning and quantum mechanics. This phase-space
approach is somewhat different from Aharonov \textit{et al.}'s weak
value approach \cite{aav_hardy}. Whether the two can be reconciled
remains to be seen.

This paper is organized as follows: Section \ref{classical} reviews
the classical filtering and smoothing equations when the system
process has jumps and the observations have Poissonian statistics, as
derived by Snyder \cite{snyder,snyder_book} and Pardoux \cite{pardoux_poisson}.
Section \ref{hybrid} generalizes such equations to the quantum regime
for smoothing of classical random processes coupled to quantum
systems. Sec.~\ref{phase_space} converts the quantum equations to
equivalent phase-space equations for discrete Wigner distributions.
Sec.~\ref{magnetometry} studies the application of the theory to
atomic magnetometry. Sec.~\ref{hardy} studies Hardy's paradox using
quantum smoothing in discrete phase space.

\section{\label{classical}Classical filtering and smoothing for Poissonian
observations}
Define $x_t$ as the classical system random process, the \textit{a
  priori} probability density of which satisfies the differential
Chapman-Kolmogorov equation \cite{gardiner}
\begin{align}
\parti{P(x,t)}{t}
&= \mathcal L_C P(x,t),
\\
\mathcal L_C P(x,t)
&\equiv -\sum_\mu \parti{}{x_\mu}\Bk{A_\mu P(x,t)}
\nonumber\\&\quad
+\frac{1}{2}\sum_{\mu,\nu}\parti{^2}{x_\mu\partial x_\nu}
\Bk{B_{\mu\nu} P(x,t)}
\nonumber\\&\quad
+\int dx' \big[J(x|x',t) P(x',t) 
- J(x'|x,t)P(x,t)\big],
\label{ck}
\end{align}
where $J(x|x',t)$ is the probability density per unit time that $x_t$
will jump from $x'$ to $x$. For an obseration process with Poissonian
noise, the conditional probability density is
\begin{align}
P(\delta N_{\mu t}|x_t) &= \exp\Bk{-\lambda_\mu(x_t,t)\delta t}
\frac{[\lambda_\mu(x_t,t)\delta t]^{\delta N_\mu}}{\delta N_\mu!},
\end{align}
with the continuous-time limit
\begin{align}
dN_{\mu t}^2 &= dN_{\mu t},
\\
P(dN_{\mu t} = 0|x_t) &= 1- \lambda_\mu(x_t,t) dt,
\\
P(dN_{\mu t} = 1|x_t) &= \lambda_\mu(x_t,t) dt.
\end{align}
Defining the observation record in the time interval $t_0 \le t < t$
as
\begin{align}
dN_{[t_0,t)} &= \BK{dN_t, t_0\le t < t}
\end{align}
and the filtering probability density as the probability density
of $x_t$ conditioned upon past observations, given by
\begin{align}
F(x,t) &\equiv P(x_t = x|dN_{[t_0,t)}),
\end{align}
the It\=o stochastic differential equation for $F(x,t)$ is
called the Snyder equation and given by
\cite{snyder,snyder_book}
\begin{align}
dF &= dt \mathcal L_C F
+\sum_\mu \bk{dN_{\mu t}-dt\Avg{\lambda_\mu}_F}
\Avg{\lambda_\mu}_F^{-1}
\nonumber\\&\quad\times
\bk{\lambda_\mu-\Avg{\lambda_\mu}_F}F,
\label{snyder}\\
\Avg{\lambda_\mu}_F &\equiv \int dx \lambda_\mu(x,t)F(x,t).
\end{align}
A linear equation for an unnormalized $F(x,t)$ was derived by Pardoux
and given by \cite{pardoux_poisson}
\begin{align}
df &= dt \mathcal L_C f
+\sum_\mu \bk{dN_{\mu t}-dt}(\lambda_\mu - 1)f,
\label{zakai_poisson}
\end{align}
with
\begin{align}
F(x,t) &= \frac{f(x,t)}{\int dx f(x,t)}.
\end{align}
To perform smoothing in the time-symmetric form
\cite{pardoux_poisson}, first solve for an unnormalized retrodictive
likelihood function $P(dN_{[t,T)}|x_t=x)\propto g(x,t)$ using the
adjoint of Eq.~(\ref{zakai_poisson}),
\begin{align}
-dg &= dt\mathcal L_C^* g + 
\sum_\mu \bk{dN_{\mu t}-dt}(\lambda_\mu - 1)g,
\label{bzakai}
\end{align}
to be solved backward in time with final condition $g(x,T)\propto 1$.
The smoothing probability density at time $\tau$ given the observation
record $dN_{[t_0,T)}$ in the time interval $t_0\le \tau \le T$ is then
\begin{align}
P(x_\tau=x|dN_{[t_0,T)}) &= \frac{g(x,\tau)f(x,\tau)}{\int dx
    g(x,\tau)f(x,\tau)}.
\label{csmooth}
\end{align}

\section{\label{hybrid}Hybrid classical-quantum filtering and
  smoothing for Poissonian observations}
Using the same approach as Refs.~\cite{tsang_prl,tsang_pra}, it is not
difficult to generalize the above classical equations to the quantum
regime for hybrid classical-quantum filtering and smoothing.  Define
$x_t$ as the classical system process that one wishes to estimate,
which is coupled to a quantum system under measurements. As before,
the quantum backaction from the quantum system to the classical one is
assumed to be negligible.  Define the hybrid density operator that
describes the joint statistics of the classical and quantum systems
\cite{hybrid} as $\hat\rho(x,t)$.  The \textit{a priori} evolution
of $\hat\rho(x_t,t)$ is governed by
\begin{align}
\parti{\hat\rho(x,t)}{t} &= 
\mathcal L\hat\rho(x,t),
\\
\mathcal L\hat\rho(x,t) &\equiv \mathcal L_0\hat\rho(x,t)
+\mathcal L_I(x)\hat\rho(x,t)
+\mathcal L_C\hat\rho(x,t),
\end{align}
where $\mathcal L_0$ is the superoperator that governs the evolution
of the quantum system, $\mathcal L_I$ is the superoperator that
describes the coupling of the classical system to the quantum system,
via an interaction Hamiltonian for example, and $\mathcal
L_{\textrm{C}}$ is the Chapman-Kolmogorov operator defined by
Eq.~(\ref{ck}). The measurement, on the other hand, is described
by the quantum Bayes theorem,
\begin{align}
\hat\rho(x_t|\delta N_{\mu t}) &= 
\frac{\hat M(\delta N_{\mu t}|x_t)\hat\rho(x_t)
\hat M^\dagger(\delta N_{\mu t}|x_t)}
{\int dx_t \trace(\textrm{numerator})},
\end{align}
where the measurement operator with Poissonian statistics is
\begin{align}
\hat M(\delta N_{\mu t}|x_t) &= 
\exp\Bk{-\frac{1}{2}\hat L_\mu^\dagger(x_t,t)\hat L_\mu(x_t,t)\delta t}
\nonumber\\&\quad\times
\frac{[\hat L_\mu (x_t,t)\sqrt{\delta t}]^{\delta N_{\mu t}}}
{\sqrt{\delta N_{\mu t}!}},
\end{align}
where $\hat L_\mu(x_t,t)$ is a hybrid operator, an annihilation
operator for example, and can also depend on $x_t$.  In the
continuous-time limit,
\begin{align}
\hat M(dN_{\mu t}=0|x_t) &= \hat 1-\frac{1}{2}
\hat L_\mu^\dagger(x_t,t)\hat L_\mu(x_t,t)dt,
\\
\hat M(dN_{\mu t}=1|x_t) &= \hat L_\mu(x_t,t)\sqrt{dt}.
\end{align}
After some algebra, the stochastic differential equation for
the filtering hybrid density operator, defined as
\begin{align}
\hat F(x,t) &\equiv \hat\rho(x_t=x|dN_{[t_0,t)}),
\end{align}
is given by \cite{gardiner_zoller,holevo}
\begin{align}
d\hat F &= dt\mathcal L\hat F
+dt\sum_\mu \bk{\hat L_\mu\hat F\hat L_\mu^\dagger
-\frac{1}{2}\hat L_\mu^\dagger\hat L_\mu\hat F
-\frac{1}{2}\hat F\hat L_\mu^\dagger\hat L_\mu}
\nonumber\\&\quad
+\sum_\mu \bk{dN_{\mu t}-dt\Avg{\hat L_\mu^\dagger\hat L_\mu}_{\hat F}}
\Avg{\hat L_\mu^\dagger\hat L_\mu}_{\hat F}^{-1}
\nonumber\\&\quad\times
\bk{\hat L_\mu\hat F\hat L_\mu^\dagger 
- \Avg{\hat L_\mu^\dagger\hat L_\mu}_{\hat F}\hat F},
\label{qsnyder}
\end{align}
where
\begin{align}
\Avg{\hat L_\mu^\dagger\hat L_\mu}_{\hat F} &\equiv
\int dx \trace\Bk{\hat L_\mu^\dagger(x,t)\hat L_\mu(x,t)\hat F(x,t)}.
\end{align}
Equation (\ref{qsnyder}) is a quantum generalization of the Snyder
equation [Eq.~(\ref{snyder})].  A linear version of Eq.~(\ref{qsnyder}),
analogous to Eq.~(\ref{zakai_poisson}), may be written as
\cite{holevo}
\begin{align}
d\hat f &= dt\mathcal L\hat f
+dt\sum_\mu \bk{\hat L_\mu\hat f\hat L_\mu^\dagger
-\frac{1}{2}\hat L_\mu^\dagger\hat L_\mu\hat f
-\frac{1}{2}\hat f\hat L_\mu^\dagger\hat L_\mu}
\nonumber\\&\quad
+\sum_\mu \bk{dN_{\mu t}-dt}
\bk{\hat L_\mu\hat f\hat L_\mu^\dagger - \hat f},
\label{qzakai}\\
\hat F(x,t) &= \frac{\hat f(x,t)}{\int dx \trace[\hat f(x,t)]}.
\end{align}
The classical incoherent limit of Eq.~(\ref{qzakai}) is obviously
Eq.~(\ref{zakai_poisson}), and Eq.~(\ref{qzakai}) can be verified
against Eq.~(\ref{qsnyder}) by normalizing the former using It\=o
rule.

To perform smoothing, one also needs to solve for the unnormalized
hybrid effect operator $\hat E(dN_{[t,T)}|x_t = x)\propto\hat g(x,t)$
using the adjoint of Eq.~(\ref{qzakai}) \cite{tsang_prl,tsang_pra},
\begin{align}
-d\hat g &= dt\mathcal L^*\hat g
+dt\sum_\mu \bk{\hat L_\mu^\dagger\hat g\hat L_\mu
-\frac{1}{2}\hat g\hat L_\mu^\dagger\hat L_\mu
-\frac{1}{2}\hat L_\mu^\dagger\hat L_\mu\hat g}
\nonumber\\&\quad
+\sum_\mu \bk{dN_{\mu t}-dt}
\bk{\hat L_\mu^\dagger \hat g\hat L_\mu - \hat g},
\end{align}
where the final condition is $\hat g(x,T)\propto \hat 1$ and the
adjoint is defined with respect to the Hilbert-Schmidt inner product
\begin{align}
\Avg{\hat g(x,t), \hat f(x,t)} &\equiv 
\int dx \trace\Bk{\hat g(x,t)\hat f(x,t)},
\\
\Avg{\hat g(x,t), \mathcal L\hat f(x,t)} &=
\Avg{\mathcal L^*\hat g(x,t), \hat f(x,t)}.
\end{align}
The smoothing probability density is then
\begin{align}
h(x,\tau) \equiv P(x_\tau=x|dN_{[t_0,T)}) &= 
\frac{\trace[\hat g(x,\tau)\hat f(x,\tau)]}
{\int dx\trace[\hat g(x,\tau)\hat f(x,\tau)]}.
\label{qsmooth}
\end{align}
Incorporating the Gaussian measurements considered in
Refs.~\cite{tsang_prl,tsang_pra} into the equations above is
straightforward. This is useful, for example, when both photon
counting and homodyne detection are performed in a quantum optics
experiment \cite{foster}.  With Poissonian observations $dN_t$ and
Gaussian observations $dy_t$, the resulting filtering equation for
$\hat F(x,t)$ is
\begin{align}
d\hat F &= dt\mathcal L\hat F
+dt\sum_\mu \bk{\hat L_\mu\hat F\hat L_\mu^\dagger
-\frac{1}{2}\hat L_\mu^\dagger\hat L_\mu\hat F
-\frac{1}{2}\hat F\hat L_\mu^\dagger\hat L_\mu}
\nonumber\\&\quad
+\frac{dt}{8}
\bk{2\hat C^T R^{-1}\hat F\hat C^\dagger - \hat C^{\dagger T}R^{-1}\hat C
\hat F - \hat F\hat C^{\dagger T}R^{-1}\hat C}
\nonumber\\&\quad
+\sum_\mu \bk{dN_{\mu t}-dt\Avg{\hat L_\mu^\dagger\hat L_\mu}_{\hat F}}
\Avg{\hat L_\mu^\dagger\hat L_\mu}_{\hat F}^{-1}
\nonumber\\&\quad\times
\bk{\hat L_\mu\hat F\hat L_\mu^\dagger 
- \Avg{\hat L_\mu^\dagger\hat L_\mu}_{\hat F}\hat F}
\nonumber\\&\quad
+\frac{1}{2}\Bk{\bk{\hat C-\avg{\hat C}_{\hat F}}^TR^{-1}
d\eta_t\hat F + \textrm{H.c.}},
\\
d\eta_t &\equiv dy_t - \frac{dt}{2}\Avg{\hat C + \hat C^\dagger}_{\hat F}, 
\end{align}
where $\hat C=\hat C(x,t)$ is a vector of hybrid operators, $R= R(t)$
is a positive-definite matrix, $d\eta_t$ is a vectoral Wiener
increment with covariance matrix $R(t)dt$, and H.~c.\ denotes
Hermitian conjugate.

The equation for $\hat f(x,t)$ is
\begin{align}
d\hat f &=dt\mathcal L\hat f
+dt\sum_\mu \bk{\hat L_\mu\hat f\hat L_\mu^\dagger
-\frac{1}{2}\hat L_\mu^\dagger\hat L_\mu\hat f
-\frac{1}{2}\hat f\hat L_\mu^\dagger\hat L_\mu}
\nonumber\\&\quad
+\frac{dt}{8}
\bk{2\hat C^T R^{-1}\hat f\hat C^\dagger - \hat C^{\dagger T}R^{-1}\hat C
\hat f - \hat f\hat C^{\dagger T}R^{-1}\hat C}
\nonumber\\&\quad
+\sum_\mu \bk{dN_{\mu t}-dt}
\bk{\hat L_\mu\hat f\hat L_\mu^\dagger - \hat f}
\nonumber\\&\quad
+\frac{1}{2}\bk{\hat C^TR^{-1}
dy_t\hat f + \textrm{H.c.}},
\label{qzakai2}
\end{align}
and for $\hat g(x,t)$,
\begin{align}
-d\hat g &=dt\mathcal L^*\hat g
+dt\sum_\mu \bk{\hat L_\mu^\dagger\hat g\hat L_\mu
-\frac{1}{2}\hat g\hat L_\mu^\dagger\hat L_\mu
-\frac{1}{2}\hat L_\mu^\dagger\hat L_\mu\hat g}
\nonumber\\&\quad
+\frac{dt}{8}
\bk{2\hat C^{\dagger T} R^{-1}\hat g\hat C  
- \hat g\hat C^{\dagger T}R^{-1}\hat C
- \hat C^{\dagger T}R^{-1}\hat C\hat g}
\nonumber\\&\quad
+\sum_\mu \bk{dN_{\mu t}-dt}
\bk{\hat L_\mu^\dagger\hat g\hat L_\mu - \hat g}
\nonumber\\&\quad
+\frac{1}{2}\bk{\hat g\hat C^TR^{-1}
dy_t + \textrm{H.c.}}.
\label{bzakai2}
\end{align}

\section{\label{phase_space}Quantum smoothing in phase space}
One method of solving Eqs.~(\ref{qsmooth}), (\ref{qzakai2}), and
(\ref{bzakai2}) for hybrid smoothing is to use Wigner distributions
\cite{tsang_prl,tsang_pra}. For a quantum system with a discrete
degree of freedom, such as spin, angular momentum, or an $N$-level
system, one may define the discrete Wigner distribution, according to
Feynman \cite{feynman} and Wootters \cite{wootters}, as
\begin{align}
f(q,p,x,t) &\equiv \frac{1}{N}\trace\Bk{\hat f(x,t)\hat W(q,p)},
\label{discrete_wigner}
\\
q &\in \BK{0,1,\dots,N-1},\\
p &\in \BK{0,1,\dots,N-1}.
\end{align}
The operator $\hat W(q,p)$ for prime $N$ is
\begin{align}
&\quad \hat W(q,p) 
\nonumber\\&\equiv
\bigg\{\begin{array}{cc}
\frac{1}{2}\Bk{(-1)^{q}\hat\sigma_z + (-1)^{p}\hat\sigma_x
+(-1)^{q+p}\hat\sigma_y + \hat 1}, &
N = 2;
\\
\sum_{q_1,q2}\delta_{2q,q_1+q_2}
\exp\Bk{\frac{2\pi i}{N}p(q_1-q_2)}\ket{q_1}\bra{q_2},
&
N> 2.
\end{array}
\label{wigner_operator}
\end{align}
$\hat\sigma_x$, $\hat\sigma_y$, and $\hat\sigma_z$ are Pauli matrices,
$\ket{q_1}$ and $\ket{q_2}$ are eigenstates of $\hat q$, and modular
arithmetic with modulus $N$ is implicitly assumed.  For a nonprime
$N$, the system can be decomposed into subsystems with prime $N$'s and
the Wigner distribution can be defined using $\hat W(q,p)$ for each
subsystem \cite{wootters}. 

An alternative definition in a $2N\times 2N$ phase space, first
suggested by Hannay and Berry \cite{hannay}, is
\begin{align}
\tilde f( q, p,x,t) &\equiv \frac{1}{2N}\trace\Bk{\hat f(x,t)
\hat w( q, p)},
\nonumber\\
 q &\in  \BK{0,\frac{1}{2},\dots,N-\frac{1}{2}},
\nonumber\\
 p &\in  \BK{0,\frac{1}{2},\dots,N-\frac{1}{2}},
\nonumber\\
\hat w( q, p) &\equiv 
\sum_{u}\exp\bk{\frac{4\pi i p u}{N}}\ket{ q+u}\bra{ q - u},
\nonumber\\
u &\in \BK{-\frac{N}{2}+\frac{1}{2}, -\frac{N}{2}+1,\dots,\frac{N}{2}},
\label{wigner_2N}
\end{align}
where the matrix elements with noninteger indices are assumed to be zero.
One may also use either Wigner function to
describe the energy level $n$ and phase $\phi$ of a harmonic
oscillator by letting $n = q$, $\phi = 2\pi p/N$, and taking the
$N\to\infty$ limit at the end of a calculation \cite{vaccaro,luks}.

Both definitions have a particularly desirable property for the
purpose of smoothing, namely,
\begin{align}
\trace\Bk{\hat g(x,t)\hat f(x,t)} &= 
N\sum_{q,p} g(q,p,x,t)f(q,p,x,t),
\\
 &= 
2N\sum_{ q, p} \tilde g( q, p,x,t)\tilde f( q, p,x,t),
\end{align}
so the smoothing probability density can be written in terms
of the Wigner distributions as
\begin{align}
h(x,\tau) &= \frac{\sum_{q,p}g(q,p,x,\tau)f(q,p,x,\tau)}{\int dx
\sum_{q,p}g(q,p,x,\tau)f(q,p,x,\tau)}
\label{smooth1}
\end{align}
or
\begin{align}
h(x,\tau) &= \frac{\sum_{ q, p}\tilde g( q, p,x,\tau)
\tilde f( q, p,x,\tau)}{\int dx
\sum_{ q, p}\tilde g( q, p,x,\tau)\tilde f( q, p,x,\tau)}.
\label{smooth2}
\end{align}
Equations (\ref{smooth1}) and (\ref{smooth2}) become equivalent to the
classical smoothing density given by Eq.~(\ref{csmooth}), with the
quantum degrees of freedom marginalized, if $f$ and $g$ or $\tilde f$
and $\tilde g$ are nonnegative and can be regarded as classical
probability densities.  The hybrid smoothing problem can then be
solved using classical filtering and smoothing techniques.

If one desires to obtain smoothing estimates of the quantum degrees of
freedom, a quantum smoothing quasiprobability distribution may be
defined as
\begin{align}
h(q,p,x,\tau) &= \frac{g(q,p,x,\tau)f(q,p,x,\tau)}
{\int dx\sum_{q,p}g(q,p,x,\tau)f(q,p,x,\tau)}
\label{smooth_qpd}
\end{align}
or
\begin{align}
\tilde h( q, p,x,\tau) &=
\frac{\tilde g( q, p,x,\tau)\tilde f( q, p,x,\tau)}
{\int dx\sum_{ q, p}\tilde g( q, p,x,\tau)
\tilde f( q, p,x,\tau)}.
\end{align}
From the perspective of estimation theory, these definitions of
quantum smoothing distributions are arguably the most natural, for
they both give the correct classical smoothing distribution when the
quantum degrees of freedom are marginalized, are equivalent to the
smoothing distributions in classical smoothing theory when $f$ and $g$
or $\tilde f$ and $\tilde g$ are nonnegative, and are explicitly
normalized.

There are many other equally qualified definitions of the Wigner
distribution in discrete or periodic phase space
\cite{gibbons,agarwal}. Choosing which definition to use depends on
the application.  The Feynman-Wootters distribution is defined only on
the eigenvalues of $\hat q$ and $\hat p$, so it appears more physical,
but the Hannay-Berry definition is easier to calculate analytically
for arbitrary $N$ and, as shown in Appendix \ref{weak_measure}, naturally
arises from the statistics of weak measurements.

\section{\label{magnetometry}Atomic Magnetometry}
\subsection{Optimal smoothing}
\begin{figure}[htbp]
\centerline{\includegraphics[width=0.48\textwidth]{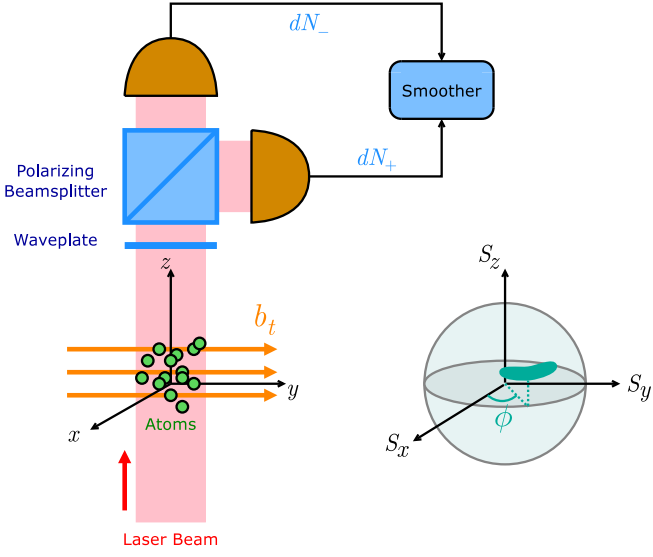}}
\caption{(color online). Left: basic setup of atomic magnetometer as
  considered in Refs.~\cite{kuzmich,geremia,stockton,bouten}. Right:
  the spherical phase space for spin.}
\label{magnetometer}
\end{figure}

An important application of quantum estimation theory is atomic
magnetometry
\cite{budker,kuzmich,thomsen,geremia,stockton,bouten,petersen}. Consider
the setup described in Refs.~\cite{geremia,stockton,bouten,kuzmich}
and depicted in Fig.~\ref{magnetometer}.  An atomic spin ensemble is
initially prepared in a coherent state with the mean collective spin
vector along the $x$ axis. Let the magnetic field be polarized along
$y$ axis and given by
\begin{align}
b_t \equiv x_{1t},
\end{align}
a component of the classical system process to be estimated. The
magnetic field introduces Larmor precession to the spin via the
interaction Hamiltonian
\begin{align}
\hat H_I(x) &= -\gamma b \hat S_y,
\\
\mathcal L_I(x)\hat F(x,t) &= -\frac{i}{\hbar}\Bk{\hat H_I(x),\hat F(x,t)}
=\frac{i\gamma}{\hbar} b\Bk{\hat S_y,\hat F},
\label{larmor}
\end{align}
where $\hat S_y$ is the $y$ component of the spin vector
operator and $\gamma$ is the gyromagnetic ratio. Under continuous
optical polarimetry measurements, the stochastic equation for the
filtering density operator $\hat F(x,t)$ has been derived by Bouten
\textit{et al.}\ \cite{bouten} and is given by
\begin{align}
d\hat F &= dt\Big\{\mathcal L_C\hat F + \frac{i\gamma}{\hbar}b\Bk{\hat S_y,\hat F}
+|a|^2
\nonumber\\&\quad\times
 \Bk{\cos(\kappa \hat m)\hat F \cos(\kappa\hat m)
+\sin(\kappa \hat m)\hat F \sin(\kappa\hat m)
-\hat F}\Big\}
\nonumber\\&\quad
+\sum_{\mu=+,-} \bk{dN_{\mu t}-dt\Avg{\hat L_\mu^\dagger\hat L_\mu}_{\hat F}}
\nonumber\\&\quad\times
\Avg{\hat L_\mu^\dagger\hat L_\mu}_{\hat F}^{-1}
\bk{\hat L_\mu\hat F\hat L_\mu^\dagger
- \Avg{\hat L_\mu^\dagger\hat L_\mu}_{\hat F}\hat F},
\label{qsnyder_magnetometry}
\end{align}
which is in the form of Eq.~(\ref{qsnyder}), with
\begin{align}
\hat m &\equiv \frac{\hat S_z}{\hbar},
&
\hat L_{\pm} &= \frac{a}{\sqrt{2}}\Bk{\cos(\kappa\hat m)\pm\sin(\kappa\hat m)},
\end{align}
$\kappa$ is the light-spin coupling parameter and $a$ is the
normalized optical envelope. The linear predictive and retrodictive
equations for $\hat f(x,t)$ and $\hat g(x,t)$ become
\begin{align}
d\hat f &= dt\Big\{\mathcal L_C\hat f + \frac{i\gamma}{\hbar} b\Bk{\hat S_y,\hat f}
+ |a|^2
\nonumber\\&\quad\times
 \Bk{\cos(\kappa \hat m)\hat f \cos(\kappa\hat m)
+\sin(\kappa \hat m)\hat f \sin(\kappa\hat m)
-\hat f}\Big\}
\nonumber\\&\quad
+\sum_{\mu=+,-} \bk{dN_{\mu t}-dt}\bk{\hat L_\mu\hat f\hat L_\mu^\dagger- \hat f},
\label{predict}\\
-d\hat g &= dt\Big\{\mathcal L_C^*\hat g - \frac{i\gamma}{\hbar}  b\Bk{\hat S_y,\hat g}
+dt |a|^2
\nonumber\\&\quad\times
 \Bk{\cos(\kappa \hat m)\hat g \cos(\kappa\hat m)
+\sin(\kappa \hat m)\hat g \sin(\kappa\hat m)
-\hat g}\Big\}
\nonumber\\&\quad
+\sum_{\mu=+,-} \bk{dN_{\mu t}-dt}\bk{\hat L_\mu^\dagger\hat g\hat L_\mu- \hat g}.
\label{retrodict}
\end{align}
After solving Eq.~(\ref{predict}) forward in time for $\hat f(x,\tau)$
and Eq.~(\ref{retrodict}) backward in time for $\hat g(x,\tau)$, the
smoothing probability distribution is given by
\begin{align}
h(x,\tau) &= \frac{\trace[\hat g(x,\tau)\hat f(x,\tau)]}
{\int dx \trace[\hat g(x,\tau)\hat f(x,\tau)]},
\end{align}
which can be used to produce the optimal estimate and the associated
error of the system process $x_\tau$, including the magnetic field
$b_\tau \equiv x_{1\tau}$.

\subsection{Smoothing in phase space}
The usual strategy of solving the quantum estimation problem is to
take the $S_x \gg S_y$, $S_x \gg S_z$ limit, assume $\hat S_y$ and
$\hat S_z$ are continuous, and approximate the conditional quantum
state as a Gaussian state
\cite{thomsen,geremia,stockton,kuzmich,petersen}. This is akin to
approximating the spherical phase space with a flat one near $\mathbf
S = (S_x,0,0)$. While the Gaussian approximation is probably the most
practical, in order to illustrate the discrete phase-space formalism
proposed in Sec.~\ref{phase_space}, I shall first attempt to convert
Eqs.~(\ref{predict}) and (\ref{retrodict}) to stochastic equations for
discrete Wigner distributions in the $2N\times2N$ phase space before
making further approximations.

Let
\begin{align}
\hat m &= \hat q-s, & N &= 2s + 1,
\end{align}
where $s$ is the total spin number. Then
\begin{align}
\hat\phi &\equiv \frac{2\pi\hat p}{N}
\end{align}
is the operator for the azimuthal angle of the spin vector.  I shall
use $m$ and $\phi$ as the phase-space variables instead of
$q$ and $p$, and let
\begin{align}
f(m,\phi) &= \tilde f\bk{q = m+s,p = \frac{N\phi}{2\pi}}.
\end{align}
First consider the measurement-induced decoherence term on the second
line of Eq.~(\ref{predict}). It can be rewritten as
\begin{align}
\mathcal L_{\textrm{M}}\hat f&\equiv \frac{1}{2}\bk{e^{i\kappa \hat q}\hat f e^{-i\kappa\hat q}
+e^{-i\kappa \hat q}\hat f e^{i\kappa\hat q}}-\hat f.
\end{align}
Using the definition of the discrete Wigner function given by
Eqs.~(\ref{wigner_2N}) and $\stackrel{\hat w}{\to}$ to denote the transform to
the $2N\times2N$ phase space, it can be shown that
\begin{align}
\mathcal L_{\textrm{M}}\hat f \stackrel{\hat w}\to
& \sum_{\phi'} J(\phi-\phi')f( m,  \phi',x,t)-f(m,\phi,x,t),
\label{qjump}
\end{align}
where
\begin{align}
J(\phi-\phi') &\equiv \frac{1}{4N}
\bigg\{\frac{\sin [N(\phi-\phi'-\kappa)]}
{\tan[(\phi-\phi'-\kappa)/2]}
\nonumber\\&\quad
\frac{\sin [N(\phi-\phi'+\kappa)]}
{\tan[(\phi-\phi'+\kappa)/2]}
\bigg\}.
\end{align}
While Eq.~(\ref{qjump}) has the appearance of the jump term in the
Chapman-Kolmogorov equation [Eq.~(\ref{ck})], $J(\phi-\phi')$, which
plays the role of a jump probability density, can become negative. In
the special case of $\kappa = \pi\mu/N$, where $\mu$ is an integer,
however, $J(\phi-\phi')$ is simplified to
\begin{align}
J(\phi-\phi') &= \frac{1}{2}\bk{\delta_{\phi-\phi',\kappa}+\delta_{\phi-\phi',-\kappa}},
\end{align}
and the measurement-induced decoherence introduces random azimuthal
jumps in steps of $\kappa$ to the spin vector around the $z$ axis. In
the limit of $N \to \infty$, $\phi$ becomes approximately continuous,
$\kappa \approx \pi\mu/N$, and Eq.~(\ref{qjump}) can be rewritten as
\begin{align}
\mathcal L_{\textrm{M}}\hat f
&\stackrel{\hat w}\to
\frac{1}{2}\Big[f(m,\phi+\kappa,x,t)
+f(m,\phi-\kappa,x,t)\Big]
\nonumber\\&\quad
-f(m,\phi,x,t).
\end{align}
The $N\to\infty$ limit is akin to approximating the spin system as a
harmonic oscillator \cite{vaccaro} and the spherical phase space as a
cylindrical one. If $\kappa \ll \avg{\Delta\hat\phi^2}^{1/2}$, we can 
further make the diffusive approximation:
\begin{align}
\mathcal L_{\textrm{M}}\hat f
&\stackrel{\hat w}\to
\frac{\kappa^2}{2}\partit{}{\phi}f(m,\phi,x,t).
\end{align}

Next, consider the Larmor precession term $(i\gamma/\hbar) b[\hat S_y,\hat f]$.
In terms of $\hat m$ and $\hat\phi$,
\begin{align}
\hat S_y &= \frac{\hbar}{2i}
\Big[\exp(-i\hat\phi)\sqrt{s(s+1)-\hat m(\hat m+1)}
\nonumber\\&\quad
-\exp(i\phi)\sqrt{s(s+1)-\hat m(\hat m-1)}\Big].
\end{align}
With this form, it is difficult to convert the Larmor precession term
to the phase-space picture analytically, so we again make the
cylindrical phase-space approximation with $s \gg \avg{\hat m},
\avg{\Delta\hat m^2}^{1/2}$, so that the spin vector distribution is
concentrated near the equator. This approximation is valid when the
magnetic field is small and fluctuating around zero, or a control,
such as an applied magnetic field \cite{thomsen,geremia,stockton} or
an adjustable direction of the optical beam, is present to realign the
spin vector with respect to the optical beam propagation direction.
Then
\begin{align}
\hat S_y &\approx - \hbar s\sin\hat\phi,
\\
\mathcal L_I\hat f &\stackrel{\hat w}{\to}
-\gamma b s\cos \phi
\Bk{f(m+\frac{1}{2},\phi,x,t)
- f(m-\frac{1}{2},\phi,x,t)}.
\end{align}
Although this looks like the jump term in Eq.~(\ref{ck}), the apparent
jump probability density is again negative. To make the classical
connection, assume that $m$ is continuous and approximate the
difference as a derivative:
\begin{align}
\mathcal L_I\hat f &\stackrel{\hat w}{\to} 
-\gamma b s\cos\phi\parti{}{m}f(m,\phi,x,t),
\end{align}
which becomes equivalent to the drift term in Eq.~(\ref{ck}) with $A_m =
\gamma b s\cos\phi$.

Finally, let us consider the terms $\hat L_\pm\hat f\hat
L_\pm^\dagger$ in Eq.~(\ref{predict}). It is not difficult to show
that, in the continuous $\phi$ limit,
\begin{align}
\hat L_\pm\hat f\hat L_\pm^\dagger &\stackrel{\hat w}{\to}
|a|^2\bigg\{\frac{1}{2} \Bk{f(m,\phi+\kappa,x,t)
+f(m,\phi-\kappa,x,t)}
\nonumber\\&\quad
\pm\sin (2\kappa m) f(m,\phi,x,t)\bigg\}.
\end{align}
These terms do not have exact analogs in the corresponding
classical equation [Eq.~(\ref{zakai_poisson})], unless we make
the $\kappa \ll \avg{\Delta\hat\phi^2}^{1/2}$ approximation,
\begin{align}
\hat L_\pm\hat f\hat L_\pm^\dagger &\stackrel{\hat w}{\to}
|a|^2\bigg\{f(m,\phi,x,t) + \frac{\kappa^2}{2}\partit{}{\phi}f(m,\phi,x,t)
\nonumber\\&\quad
\pm\sin (2\kappa m) f(m,\phi,x,t)\bigg\}
\\
&\approx
|a|^2\Bk{1\pm\sin (2\kappa m)} f(m,\phi,x,t).
\end{align}
Summarizing, a classical model of atomic magnetometry can be obtained
if we approximate the spherical phase space as a cylindrical one near
the equator, assume $m$ is continuous, and let $\kappa \ll
\avg{\Delta\hat\phi^2}^{1/2}$. The resulting equations
for $f(m,\phi,x,t)$ and $g(m,\phi,x,t)$ are
\begin{align}
df &= dt \bk{\mathcal L_C f -\gamma b s\cos\phi\parti{f}{m}
+\frac{|a|^2\kappa^2}{2}\partit{f}{\phi}}
\nonumber\\&\quad
+\sum_{\mu=+,-} (dN_\mu - dt)(\lambda_\mu -1) f,
\\
-dg &= dt \bk{\mathcal L_C^* g +\gamma b s\cos\phi\parti{g}{m}
+\frac{|a|^2\kappa^2}{2}\partit{g}{\phi}}
\nonumber\\&\quad
+\sum_{\mu=+,-} (dN_\mu - dt)(\lambda_\mu -1) g,
\\
\lambda_{\pm} &= |a|^2\Bk{1\pm \sin(2\kappa m)}.
\end{align}
The equivalent system equations are then
\begin{align}
dx_t &= A(x_t,t)dt + B(x_t,t)dW_t,
\\
dm_t &= dt\gamma b_t s\cos\phi_t,
\end{align}
where $dW_t$ and $d\phi_t$ are independent Wiener increments with
$dW_tdW_t^T = Q(t)dt$ and $d\phi_t^2 = |a|^2\kappa dt$. Unlike the
Gaussian model \cite{geremia,stockton,petersen}, this slightly more general
model shows that the $z$ component of the spin is coupled to $\phi_t$
via Larmor precession, as one would expect from classical dynamics,
since $S_x \approx \hbar s\cos\phi$ when $s \gg m$. The diffusion of
$\phi$ would therefore reduce the estimation accuracy in the long run.

To make the Gaussian approximation, let $\avg{\hat
  \phi_t},\avg{\Delta\hat\phi_t^2}^{1/2} \ll 1$, so that $\cos\phi_t
\approx 1$, and let $x_t$ be a Gaussian random process, such as the
Ornstein-Uhlenbeck process \cite{stockton,petersen}. If
$\kappa\avg{\hat m}, \kappa\avg{\Delta\hat m^2}^{1/2} \ll 1$, and the
effective noise covariances are $\avg{\lambda_{\pm}} \approx |a|^2$,
one can use the linear Mayne-Fraser-Potter smoother
\cite{tsang_pra,mayne,tsang_pra2}, which combines the estimates and covariances
from a predictive Kalman filter and a retrodictive Kalman filter, to
produce the optimal estimate of $x_t$. Other equivalent linear
smoothers may also be used \cite{tsang_pra2}.

\section{\label{hardy}Hardy's paradox in phase space}
In this section, I shall study Hardy's paradox \cite{hardy} in phase
space using the quantum smoothing quasiprobability distribution
defined by Eq.~(\ref{smooth_qpd}), which allows one to estimate
quantum degrees of freedom given past and future observations in a way
closely resembling classical estimation theory. The more physical and
intuitive Feynman-Wootters distribution is used, since its elements
all correspond to actual paths in the setup. I shall show that the
paradox arises because the predictive Wigner distribution becomes
negative, and quantum mechanics becomes incompatible with classical
estimation as a result. This approach is somewhat different from
Aharonov \textit{et al.}'s attempt to explain Hardy's paradox using
weak values \cite{aav_hardy}.

\begin{figure}[htbp]
\centerline{\includegraphics[width=0.45\textwidth]{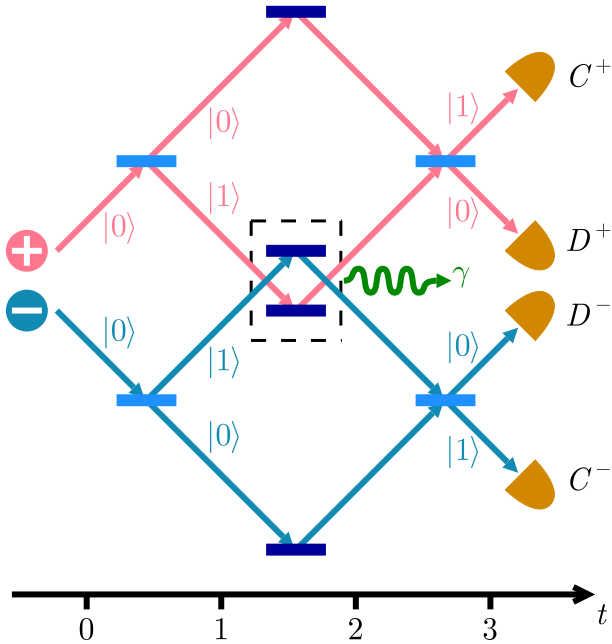}}
\caption{(Color online). Setup of Hardy's paradox.}
\label{hardy_figure}
\end{figure}

As a brief review of the paradox, consider two Mach-Zehnder
interferometers, one for a positron and one for an electron, depicted
in Fig.~\ref{hardy_figure}.  If the interferometers are physically
separate, then the setup can be configured so that the particles
always arrive at the $C^{+}$ and $C^{-}$ detectors, respectively. Now
let's make one arm of an interferometer to overlap with an arm of the
other. After the first pair of beamsplitters, the two particles may
meet in the overlapping arms, in which case they annihilate each other
with probability $1$.  With this overlapping setup, there is a $1/16$
probability that the particles will arrive at $D^{+}$ and $D^{-}$,
respectively, according to quantum theory.

The paradox arises when one tries to use classical reasoning to
estimate which arms the particles went through. If $D^{+}$ detects a
positron, then the electron must have been in the overlapping arm to
somehow influence the positron to go to $D^{+}$ instead of
$C^{+}$. The same reasoning can be applied when $D^{-}$ detects an
electron, which should mean that the positron was in the overlapping
arm. But if both particles went through the overlapping arms, they
should have annihilated each other and would not have been detected.

Denote the position of a particle in one arm as $0$ and that in the
other arm as $1$, as shown in Fig.~\ref{hardy_figure}. At the time
instant labeled $0$,
\begin{align}
\ket{\Psi}_0 &= \ket{0,0},
\end{align}
where the first number in the ket denotes the position of the
positron, the second number denotes that of the electron, and the
subscript is the time label.  The corresponding two-particle Wigner
distribution using Eqs.~(\ref{discrete_wigner}) and
(\ref{wigner_operator}) is
\begin{align}
&\quad f_0(q^{+},q^{-},p^{+},p^{-})\nonumber\\
&= \bk{\begin{array}{cccc}
f_0(0,0,0,0) & f_0(0,0,0,1) & f_0(0,0,1,0) & f_0(0,0,1,1)
\\
f_0(0,1,0,0) & f_0(0,1,0,1) & f_0(0,1,1,0) & f_0(0,1,1,1)
\\
f_0(1,0,0,0) & f_0(1,0,0,1) & f_0(1,0,1,0) & f_0(1,0,1,1)
\\
f_0(1,1,0,0) & f_0(1,1,0,1) & f_0(1,1,1,0) & f_0(1,1,1,1)
\end{array}}
\\
&=\frac{1}{4}\bk{\begin{array}{cccc}
1 & 1 & 1 & 1 \\
0 & 0 & 0 & 0 \\
0 & 0 & 0 & 0 \\
0 & 0 & 0 & 0 \end{array}}.
\end{align}
After the first pair of beamsplitters,
\begin{align}
\ket{\Psi}_1 &= \frac{1}{2}\bk{\ket{0,0}+\ket{0,1}+\ket{1,0}+\ket{1,1}},
\\
f_1(q^{+},q^{-},p^{+},p^{-}) &= 
\frac{1}{4}\bk{\begin{array}{cccc}
1 & 0 & 0 & 0 \\
1 & 0 & 0 & 0 \\
1 & 0 & 0 & 0 \\
1 & 0 & 0 & 0 \end{array}}.
\end{align}
If the annihilation did \textit{not} occur, the \textit{a posteriori}
quantum state is
\begin{align}
\ket{\Psi}_2 &= \frac{1}{\sqrt{3}}
\bk{\ket{0,0} + \ket{0,1} + \ket{1,0}},
\\
f_2(q^{+},q^{-},p^{+},p^{-}) &= \frac{1}{12}\bk{\begin{array}{cccc}
4 & 0 & 0 & 0\\
2 & 0 & 2 & 0\\
2 & 2 & 0 & 0\\
1 & -1 & -1 & 1
\end{array}}.
\end{align}
The Wigner distribution has negative elements and can no longer be
regarded as a classical phase-space probability distribution. The
negative elements, as one shall see later, can be regarded as the
culprits that cause the paradox. The predictive marginal distributions
are still nonnegative, however.  In particular,
\begin{align}
f_2(q^{+},q^{-}) &\equiv 
\sum_{p^{+},p^{-}}f_2(q^{+},q^{-},p^{+},p^{-}) \\
&=
\bk{\begin{array}{c}
f_2(0,0)\\f_2(0,1)\\f_2(1,0)\\f_2(1,1)\end{array}}
= \frac{1}{3}\bk{\begin{array}{c}
1\\ 1\\ 1\\ 0\end{array}},
\end{align}
which correctly predicts the \textit{a posteriori} position
probability distribution if one measures the positions of the
particles at that instant using strong measurements. Most importantly,
$f_2(1,1) = 0$, and the probability that one measures both particles
in the overlapping arms with strong measurements is zero.

Now perform retrodiction. Given that $D^{+}$ and $D^{-}$ click, it can
be shown that
\begin{align}
g_2(q^{+},q^{-},p^{+},p^{-}) &= \frac{1}{4}
\bk{\begin{array}{cccc}
0 & 0 & 0 & 1\\
0 & 0 & 0 & 1\\
0 & 0 & 0 & 1\\
0 & 0 & 0 & 1
\end{array}}.
\end{align}
The smoothing quasiprobability distribution at time instant $2$ 
becomes
\begin{align}
h_2(q^{+},q^{-},p^{+},p^{-}) &\propto 
f_2(q^{+},q^{-},p^{+},p^{-})g_2(q^{+},q^{-},p^{+},p^{-})
\label{h2_def}\\
&=\bk{\begin{array}{cccc}
0 & 0 & 0 & 0\\
0 & 0 & 0 & 0\\
0 & 0 & 0 & 0\\
0 & 0 & 0 & 1\end{array}},
\\
h_2(q^{+},q^{-}) &=\bk{\begin{array}{c}
0\\ 0\\ 0\\ 1
\end{array}}.
\end{align}
Hence, given that the annihilation did not occur and $D^{+}$ and
$D^{-}$ click, both particles ``reappear'' in the overlapping arms
according to quantum smoothing. This result is consistent with the
classical logic that leads one to the same paradoxical conclusion.
Mathematically, the paradox arises because the filtering estimation
according to $f_2(q^{+},q^{-})$ contradicts the smoothing estimation
according to $h_2(q^{+},q^{-})$, with the former ascertaining that the
particles cannot both be in the overlapping arms, while the latter
insisting the opposite.

To see why this cannot happen in classical estimation theory, assume
for the time being that $f_2(q^{+},q^{-},p^{+},p^{-})$ is
nonnegative. Then $f_2(1,1)$ is zero if and only if
$f_2(1,1,p^{+},p^{-})$ is zero for all $p^{+}$ and $p^{-}$. If
$f_2(1,1,p^{+},p^{-})$ is zero, so are $h_2(1,1,p^{+},p^{-})$ and
$h_2(1,1)$ according to Eq.~(\ref{h2_def}). In other words, in
classical estimation, if filtering estimates that the two particles
cannot both be in the overlapping arms, then no amount of smoothing
afterwards can alter the certainty of this fact.

Quantum smoothing, on the other hand, is able to overrule quantum
filtering because some elements of $f_2(1,1,p^{+},p^{-})$ are
negative.  This way $f_2(1,1)$ can still be zero with nonzero
$f_2(1,1,p^{+},p^{-})$ elements, and $h_2(1,1,p^{-},p^{+})$ and
$h_2(1,1)$, conditioned upon the detection outcomes, can become
nonzero. The negative elements of $f_2(q^{+},q^{-},p^{+},p^{-})$ thus
cause filtering and smoothing to produce contradictory trajectories.

\begin{figure}[htbp]
\centerline{\includegraphics[width=0.48\textwidth]{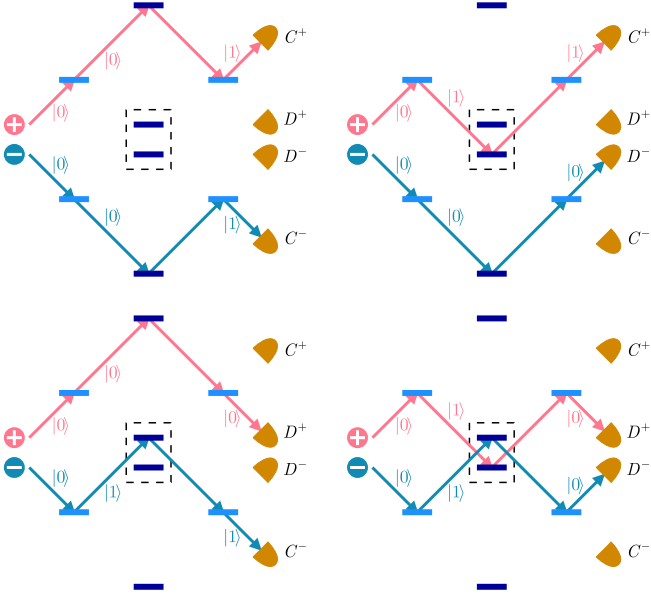}}
\caption{(color online). The most likely paths undertaken by the
  particles indicated by quantum smoothing given the detection
  outcomes, provided that annihilation did not occur. These paths
  coincide with those suggested by qualitative classical
  reasoning. When $D^+$ and $D^-$ click, the estimated paths, as shown
  in the bottom-right figure, contradict with the fact that
  annihilation did not occur and the two particles could not have both
  been in the overlapping arms.}
\label{classical_hardy}
\end{figure}

If the detection outcomes are different, say, $C^{+}$ and $D^{-}$
click, then
\begin{align}
g_2(q^{+},q^{-},p^{+},p^{-}) &= \frac{1}{4}
\bk{\begin{array}{cccc}0 & 1 & 0 & 0\\
0 & 1 & 0 & 0\\
0 & 1 & 0 & 0\\
0 & 1 & 0 & 0
\end{array}},
\\
h_2(q^{+},q^{-},p^{+},p^{-}) &= 
\bk{\begin{array}{cccc}
0 & 0 & 0 & 0\\
0 & 0 & 0 & 0\\
0 & 2 & 0 & 0\\
0 & -1 & 0 & 0
\end{array}},
\\
h_2(q^{+},q^{-}) &= 
\bk{\begin{array}{c}
0 \\ 0 \\ 2 \\ -1
\end{array}},
\end{align}
and we have a negative ``probability.'' Leaving aside the question of
interpreting a negative probability \cite{feynman}, $h_2(q^{+},q^{-})$
still suggests that the most likely positions are $(q^{+},q^{-}) =
(1,0)$, which are consistent with classical reasoning and do not
contradict the filtering results indicated by
$f_2(q^{+},q^{-})$. Similarly, when $C^{+}$ and $C^{-}$ click, the
most likely $(q^{+},q^{-})$ according to $h_2(q^{-},q^{+})$ is
$(0,0)$, which is again what one would expect from a classical
argument. In this example at least, the most likely positions
suggested by quantum smoothing coincide with the ones obtained by
qualitative classical reasoning, as depicted in
Fig.~\ref{classical_hardy}.

To summarize, quantum phase-space filtering and smoothing are able to
reproduce mathematically the salient features of Hardy's paradox and
identify the negativity of $f_2(q^+,q^-,p^+,p^-)$ as the culprit that
makes the classical phase-space picture and quantum theory
incompatible.


\section{Conclusion}
In conclusion, the time-symmetric smoothing theory is extended to
account for discrete variables in classical systems, quantum systems,
and observations. To illustrate the extended theory, atomic
magnetometry and Hardy's paradox are studied using quantum phase-space
smoothing. The generalized smoothing theory outlined in this paper is
expected to be useful in future quantum sensing and information
processing applications.

\section*{Acknowledgments}
Discussions with Seth Lloyd, Jeffrey Shapiro, and Yutaka Shikano are
gratefully acknowledged. This work was financially supported by the
Keck Foundation Center for Extreme Quantum Information Theory.

\appendix
\section{\label{weak_measure} Obtaining the quantum smoothing distribution by
weak measurements}
In the case of continuous variables, the quantum smoothing
distribution can be obtained from the statistics of weak position and
momentum measurements, conditioned upon past and future observations
\cite{tsang_pra}. One may also apply a similar method to the
discrete-variable case. Interestingly, the statistics of weak
measurements naturally lead to a $2N\times 2N$ phase space.

Consider consecutive $q$ and $p$ measurements of a quantum system. Let
the measurement operators be
\begin{align}
\hat M(y_q) &=\sqrt{\mathcal C} \sum_{q=0}^{N-1}
\exp\Bk{\frac{\epsilon}{2}\cos\frac{2\pi}{N}(y_q-q)}\ket{q}\bra{q},
\\
\hat M(y_p) &=\sqrt{\mathcal C} \sum_{p=0}^{N-1}
\exp\Bk{\frac{\epsilon}{2}\cos\frac{2\pi}{N}(y_p-p)}\ket{p}\bra{p},
\end{align}
where $\mathcal C$ is a normalization constant and $\epsilon$
parameterizes the measurement strength and accuracy. The probability
distribution of $y_q$ and $y_p$, conditioned upon past and future
observations, is
\begin{align}
&\quad P(y_q,y_p)
\nonumber\\
&= \frac{\trace[\hat g\hat M(y_p)\hat M(y_q)\hat f\hat M^\dagger(y_q)
\hat M^\dagger(y_p)]}{\trace(\hat g\hat f)}
\nonumber\\
&=\frac{\mathcal C^2}{N\trace(\hat g\hat f)}
\sum_{q,q',p,p'} \exp \Bk{\frac{\epsilon}{2}\cos\frac{2\pi(y_q-q)}{N}}
\nonumber\\&\quad\times
\exp\Bk{\frac{\epsilon}{2}\cos\frac{2\pi(y_q-q')}{N}}
\nonumber\\&\quad\times
\exp\Bk{\frac{\epsilon}{2}\cos\frac{2\pi}{N}(y_p-p)
+\frac{\epsilon}{2}\cos\frac{2\pi}{N}(y_p-p')}
\nonumber\\&\quad\times
\exp\Bk{\frac{2\pi i (p'q'-pq)}{N}}
\bra{p'}\hat g\ket{p}\bra{q}\hat f\ket{q'}.
\end{align}
Let
\begin{align}
\bar q &= \frac{q + q'}{2}, & u &= \frac{q'-q}{2},
& \bar p &= \frac{p + p'}{2}, & v &= \frac{p'-p}{2}.
\end{align}
Applying trigonometric identities, one obtains
\begin{align}
&\quad P(y_q,y_p)
\nonumber\\
&=\frac{\mathcal C^2}{N\trace(\hat g\hat f)}
\sum_{q,q',p,p'} \exp \Bk{\epsilon\cos\frac{2\pi(y_q-\bar q)}{N}}
\nonumber\\&\quad\times
\exp\Bk{\epsilon\cos\frac{2\pi(y_p-\bar p)}{N}}
\nonumber\\&\quad\times
\exp \Bk{-2\epsilon\cos\frac{2\pi(y_q-\bar q)}{N}\sin^2 \frac{\pi u}{N}}
\nonumber\\&\quad\times
\exp\Bk{-2\epsilon\cos\frac{2\pi(y_p-\bar p)}{N}\sin^2 \frac{\pi v}{N}}
\nonumber\\&\quad\times
\exp\Bk{\frac{4\pi i (v\bar q + \bar p u)}{N}}
\bra{\bar p + v}\hat g\ket{\bar p - v}\bra{\bar q -u}\hat f\ket{\bar q +u}.
\end{align}
Utilizing the periodic nature of the above expression, one can change
the sum in terms of $(q,q')$ to a sum in terms of $(\bar q,u)$,
\begin{align}
\sum_{q=0}^{N-1}\sum_{q'=0}^{N-1} &\to \frac{1}{2}\sum_{\bar q,u},
\\
\bar q &\in \BK{0, \frac{1}{2},\dots, N-\frac{1}{2}},
\\
u &\in \BK{-\frac{N}{2}+\frac{1}{2}, \frac{N}{2}+1,\dots,\frac{N}{2}},
\end{align}
likewise for $(p,p')$ and $(\bar p, v)$, and the matrix elements
$\bra{\bar p + v}\hat f\ket{\bar p-v}$ and $\bra{\bar q - u}\hat
g\ket{\bar q+u}$ are assumed to be zero whenever $\bar p+ v$, $\bar p
-v$, $\bar q-u$, or $\bar q+ u$ is not an integer. Thus,
\begin{align}
&\quad P(y_q,y_p)
\nonumber\\
&=\frac{N\mathcal C^2}{\trace(\hat g\hat f)}
\sum_{\bar q, \bar p} \exp \Bk{\epsilon\cos\frac{2\pi(y_q-\bar q)}{N}
+\epsilon\cos\frac{2\pi(y_p-\bar p)}{N}}
\nonumber\\&\quad\times
\tilde g(\bar q,\bar p) \tilde f(\bar q, \bar p),
\label{convolve}
\end{align}
where
\begin{align}
\tilde f(\bar q,\bar p) &= 
\frac{1}{2N}\sum_{v}
\exp\Bk{-2\epsilon\cos\frac{2\pi(y_q- \bar q)}{N}\sin^2 \frac{\pi u}{N}}
\nonumber\\&\quad\times
\exp\bk{\frac{4\pi i \bar p u }{N}}\bra{\bar q -u}\hat f\ket{\bar q+u},
\\
\tilde g(\bar q,\bar p) &= 
\frac{1}{2N}\sum_{u}
\exp \Bk{-2\epsilon\cos\frac{2\pi(y_p-\bar p)}{N}\sin^2 \frac{\pi v}{N}}
\nonumber\\&\quad\times
\exp\bk{\frac{4\pi i v\bar q}{N}}
\bra{\bar p + v}\hat g\ket{ \bar p - v}.
\end{align}
In the limit of infinitesimally weak measurements and $\epsilon \ll 1$,
\begin{align}
\tilde f(\bar q,\bar p) &\approx 
\frac{1}{2N}\sum_{v}
\exp\bk{\frac{4\pi i \bar pu}{N}}\bra{\bar q-u}\hat f\ket{\bar q +u},
\\
\tilde g(\bar q,\bar p) &\approx
\frac{1}{2N}\sum_{u}
\exp\bk{\frac{4\pi iv\bar q }{N}}
\bra{\bar p + v}\hat g\ket{\bar p - v},
\end{align}
which are precisely the discrete Wigner distributions in the $2N\times
2N$ phase space. Equation (\ref{convolve}) becomes
\begin{align}
&\quad P(y_q,y_p)
\nonumber\\
&=\mathcal C^2
\sum_{\bar q, \bar p} \exp \Bk{\epsilon\cos\frac{2\pi(y_q-\bar q)}{N}
+\epsilon\cos\frac{2\pi(y_p-\bar p)}{N}}
\tilde h(\bar q,\bar p),
\label{convolve2}
\end{align}
and can be regarded, from the perspective of classical probability
theory, as the probability distribution for noisy $q$ and $p$
measurements, when the system has a phase-space distribution given by
the quantum smoothing distribution $\tilde h(\bar q,\bar p)$. $\tilde
h(\bar q,\bar p)$ can therefore be obtained in an experiment with
small $\epsilon$ by measuring $P(y_q,y_p)$ for the same $\hat g$ and
$\hat f$ and deconvolving Eq.~(\ref{convolve2}).

\end{document}